\documentclass[slac_one]{revtex4}
\usepackage{graphicx}
\usepackage{fancyhdr}
\begin{document}

\title{Fits of the Electroweak Standard Model and Beyond using Gfitter} 

%

\author{J. Haller (for the Gfitter Group\footnote{
H.~Fl\"acher (CERN), 
M.~Goebel (Universit\"at Hamburg/DESY), 
J.~H., 
A.~H\"ocker (CERN), 
K.~M\"onig (DESY), 
J.~Stelzer (DESY)})}
\affiliation{Institut f\"ur Experimentalphysik, Universit\"at Hamburg, Hamburg, Germany}

\begin{abstract}
The global fit of the Standard Model to electroweak precision data, routinely performed
by the LEP electroweak working groups and others, has been revisited  
in view of ({\it i}) the development of the new generic fitting 
package, {\em Gfitter}, ({\it ii}) the insertion of constraints from direct 
Higgs searches at LEP and Tevatron, and ({\it iii})
a more thorough statistical interpretation of the results. 
This paper describes the {\em Gfitter} project, and presents
state-of-the-art results for the global electroweak fit in the 
Standard Model, and for a model with an extended Higgs sector.
Example results are an
estimation of the mass of the Higgs boson ($M_H\ =\ 116.4^{\,+18.3}_{\,-1.3}\,{\rm GeV}$)
and a forth-order result for the strong coupling strength 
($\alpha_S(M_Z^2)=0.1193^{\,+0.0028}_{\,-0.0027}{\rm (exp)} \pm 0.0001{\rm (theo)}$). 
Using toy Monte Carlo techniques the $p$-value of the SM has been determined ($p=0.22$).
As an example of a New Physics model constraints are derived for the Two Higgs Doublet Model 
of Type-II using observables from the $B$ and $K$ physics sectors.
\end{abstract}

\maketitle

\thispagestyle{fancy}

\section{INTRODUCTION} 
Precision measurements allow us to probe physics at much higher energy scales than the 
masses of the particles directly involved in experimental reactions by exploiting 
contributions from quantum loops. A prominent example is the global fit of the 
Standard Model (SM) to electroweak precision data, routinely performed by the LEP electroweak 
working group and others (for latest results see~\cite{renton}), which demonstrated 
impressively the predictive power of electroweak unification and quantum loop 
corrections. Several theoretical libraries within and beyond the SM have been developed 
in the past, which allowed to constrain the unbound parameters of the SM and models of 
New Physics. However, most of these programs are relatively old, were implemented in 
outdated programming languages, and are difficult to maintain in line with the 
theoretical and experimental progress expected during the forthcoming era of the LHC.
These considerations led to the development of the generic fitting package 
{\em Gfitter}~\cite{gfitterweb, gfitter}, designed to provide a modular framework for 
complex fitting tasks in high-energy physics, like model testing and parameter 
estimation problems. {\em Gfitter} is implemented in C++ and relies on ROOT 
functionality. It consists of a core package providing the tools for data handling, 
fitting and statistical analyses and allows a consistent treatment of statistical, 
systematic and theoretical errors, possible correlations, and inter-parameter 
dependencies. Tools provided for statistical analyses include e.g. parameter scans, 
contours, Monte Carlo (MC) toy analyses and goodness-of-fit $p$-value evaluation. More 
details on the framework can be found at~\cite{gfitterweb}. {\em Gfitter} performs 
the minimisation of a $\chi^2$ test statistics quantifying the deviation of the experimental 
data from the predictions in a certain physics model. The theoretical calculations 
are implemented via plug-in 
libraries for the {\em Gfitter} framework. In this paper we report results which are
obtained using the first libraries implemented in the {\em Gfitter} package: SM 
predictions of the electroweak precision observables and predictions of  $B$ and 
$K$ physics observables in a model with two Higgs doublets (2HDM).

\section{THE GLOBAL ELECTROWEAK FIT}

In the global electroweak fit with {\em Gfitter} state-of-the-art calculations are 
compared with the most recent experimental data to constraint the free parameters 
of the fit and to test the goodness-of-fit. The SM parameters relevant for the global 
electroweak analysis are the coupling constants of the electromagnetic, weak and strong 
interactions, and the masses of the elementary bosons and fermions. Electroweak unification 
and simplifications arising from fixing parameters with insignificant uncertainties 
compared to the sensitivity of the fit allow to reduce the number of free fit parameters. 
The remaining free parameters are the coupling parameters 
$\Delta\alpha_{\rm had}^{(5)}(M_Z^2)$ and $\alpha_S(M_Z^2)$, the masses 
$M_Z$, $\overline{m}_c$, $\overline{m}_b$, $m_t$ and $M_H$. In addition, four free 
parameters enter to include the theoretical uncertainties of 
$M_W$, $\sin^2\theta_{\rm eff}^{l}$ and the electroweak form factors $\rho_Z^f$ and 
$\kappa_Z^f$.

For the prediction of the electroweak precision observables as measured by the LEP, SLC 
and Tevatron experiments the most up-to-date calculations are implemented in the 
{\em Gfitter} SM library using the OMS scheme. Wherever possible the results have been 
cross-checked against the ZFITTER package~\cite{zfitter}. The full two-loop and leading 
beyond-two-loop correction are available for the computation of $M_W$ and 
$\sin^2\theta_{\rm eff}^{l}$~\cite{awramik1, awramik2}. The partial and total widths 
of the $Z$ are known to leading order, while for the second order only the leading 
$m_t^2$ corrections are available. Among the new developments included is the NNNLO 
perturbative calculation of the massless QCD Adler function~\cite{adler}, contributing 
to the vector and axial-vector radiator functions in the prediction of the $Z$ hadronic 
width (and other observables). It allows to fit the strong coupling constant with 
unique theoretical accuracy. More details on the theoretical computations in 
{\em Gfitter} can be found in~\cite{gfitter}.

\begin{figure}[t]
   \includegraphics[width=88mm]{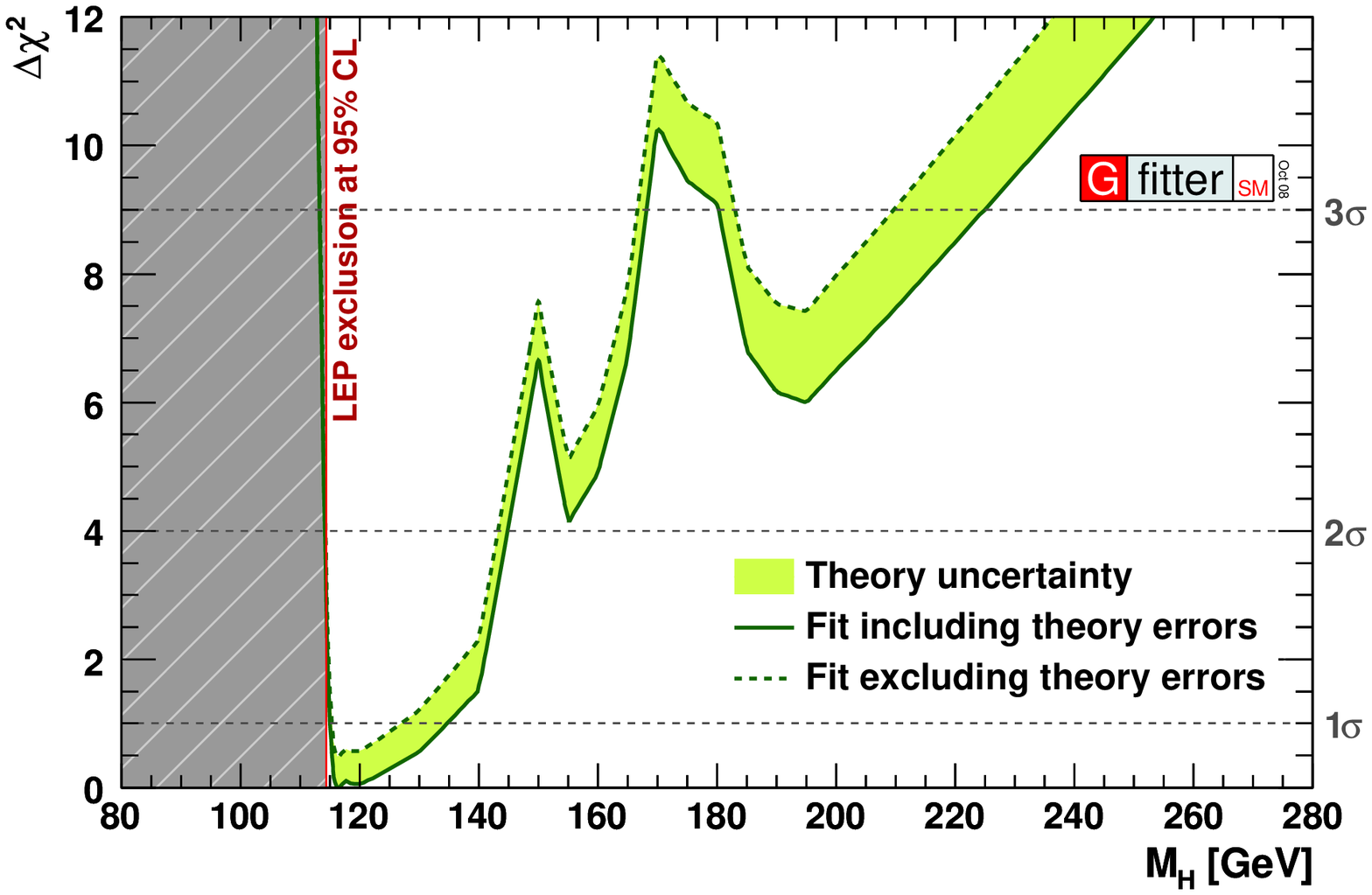}
   \includegraphics[width=88mm]{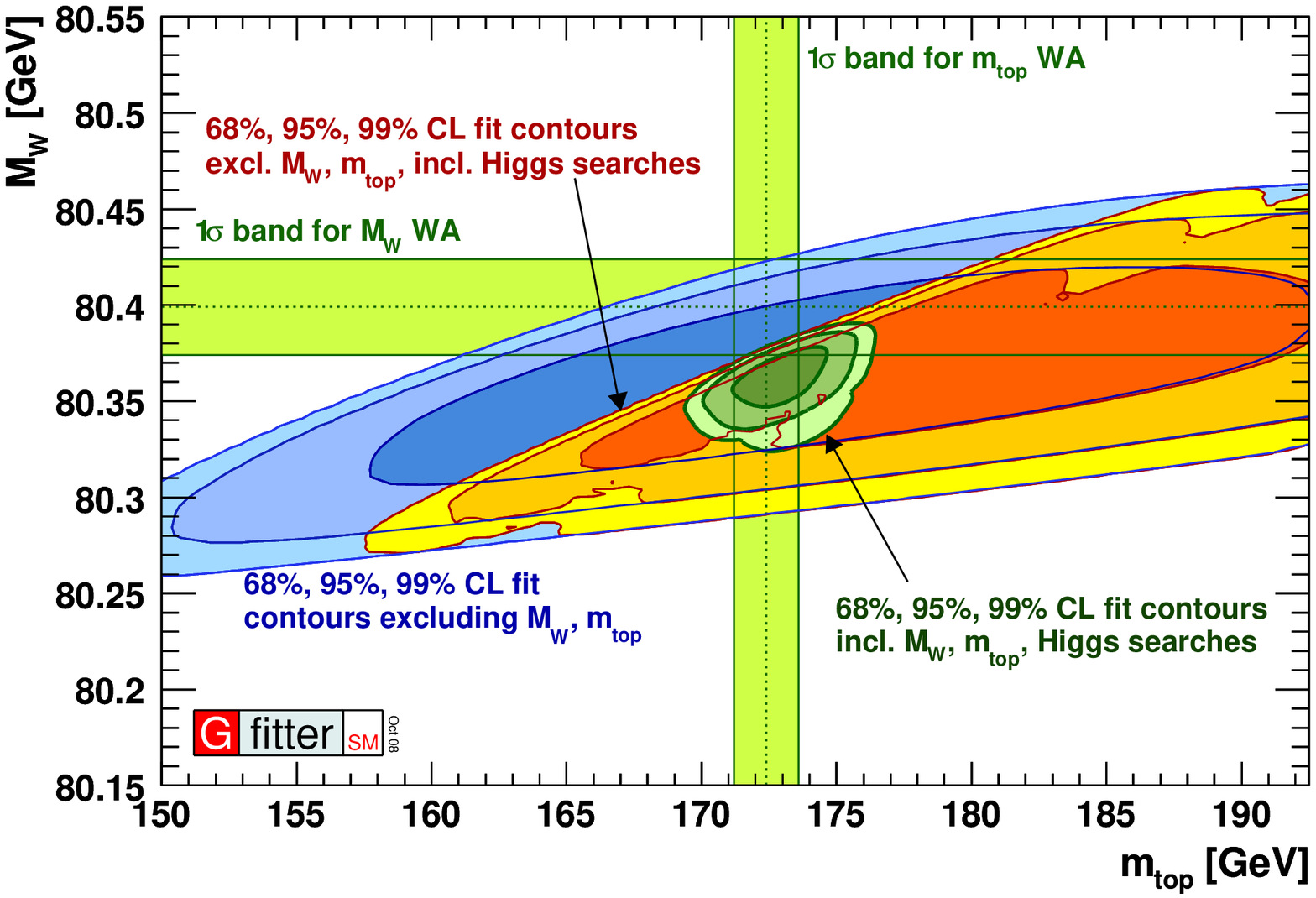}
   \caption{Left: $\Delta \chi^2$ as a function of $M_H$ for the {\em complete fit}. 
     The solid (dashed) lines give the results when including (ignoring) theoretical 
     errors. The minimum $\Delta \chi^2$ of the fit including theoretical errors is 
     used for both curves in each plot to obtain the offset-corrected $\Delta \chi^2$;
     Right: Contours of 68\%, 95\% and 99\%~CL obtained from scans of fits with fixed
     variable pairs $M_W$ vs. $m_t$ for three sets of fits explained in the main text. 
     The horizontal bands indicate the $1\sigma$ regions of measurements (world averages).
   }
   \label{fig:chi2lep}
\end{figure}

The following experimental measurements are used: The mass and width of the $Z$ boson, 
the hadronic pole cross section $\sigma^0_{\rm had}$, the partial widths ratio $R^0_\ell$, 
and the forward-backward asymmetries for leptons $A_{\rm FB}^{0,\ell}$, have been determined 
by fits to the $Z$ line-shape measured precisely at LEP (see~\cite{zsummary} and references 
therein). Measurements of the $\tau$ polarisation at LEP~\cite{zsummary} and the left-right 
asymmetry at SLC~\cite{zsummary} have been used to determine the lepton asymmetry parameter 
$A_\ell$. The corresponding $c$ and $b$-quark asymmetry parameters $A_{c(b)}$, the 
forward-backward asymmetries $A_{\rm FB}^{0,c(b)}$, and the widths ratios $R^0_c$ and $R^0_b$, 
have been measured at LEP and SLC~\cite{zsummary}. In addition, the forward-backward 
charge asymmetry measurement in inclusive hadronic events at LEP was used to directly 
determine $\sin^2\theta_{\rm eff}^{\ell}$~\cite{zsummary}. For the running quark masses 
$\overline{m}_c$ and $\overline{m}_b$ the world average values are used. For 
$\Delta\alpha_{\rm had}^{(5)}(M_Z^2)$ we use the most recent phenomenological 
result~\cite{hagiwara}. Results presented in this paper are obtained using the combined 
LEP and Tevatron results on the mass and the width of the $W$ boson~\cite{newwmass}, 
$M_{W}= (80.399\pm0.025)\,\rm GeV$, $\Gamma_{W}=(2.098\pm0.048)\,\rm GeV$, and the latest 
combined result on the top mass~\cite{newtopmass}, $m_t=(172.4\pm1.2)\,\rm GeV$, presented 
at this conference.

The direct searches for the SM Higgs Boson at LEP~\cite{higgsLEP} and the most recent 
results from the Tevatron~\cite{higgstev1, higgstev2}, leading to a 95\% confidence 
level (CL) exclusion for $M_H<114.4\,\rm GeV$ and at $M_H=170\,\rm GeV$ respectively, 
are included using a Gaussian approach that quantifies the difference between the 
observed test statistics (the log-likelihood ratios) and the expected values 
for the s+b hypothesis using the published values of the respective confidence level 
(${\rm CL}_{\rm S+B}$). A contribution to the $\chi^2$ estimator of the fit is derived for 
each Higgs mass. 
We perform global fits in two versions: the {\em standard (``blue-band'') fit} makes
use of all the available information except for the direct Higgs searches; 
the {\it complete fit} uses also the constraints from the 
direct Higgs searches.

Due to the restricted space available only example results of the electroweak fit are 
reported in the following. More fit results and thorough studies of its statistical 
properties are given in~\cite{gfitter} where also the perspectives for the LHC, 
ILC and GigaZ data are discussed. The {\em standard} ({\em complete}) {\em fit} 
converges at the global minimum value $\chi^2_{\rm min}=16.4$ 
($\chi^2_{\rm min}=18.0$) for 13 (14) degrees of freedom. The estimation for $M_H$ 
from the {\em standard fit} without the direct Higgs searches is 
$M_H\ =\ 80^{\,+30}_{\,-23}\:\rm GeV$ and the $2\sigma$ and $3\sigma$ intervals are 
respectively $[39,\,156]\,\rm GeV$ and $[26,\,209]\,\rm GeV$. The {\em complete fit} 
represents the most accurate estimation of $M_H$ considering all available data. We 
find $M_H\ =\ 116.4^{\,+18.3}_{\,-1.3}\, \rm GeV$. The resulting $\Delta\chi^2$ curve 
versus $M_H$ is shown in Fig.~\ref{fig:chi2lep} (left). The shaded band indicates the 
influence of theoretical uncertainties. The one, two and three standard deviations 
from the minimum are indicated by the crossings with the corresponding horizontal lines. 
The $2\sigma$ and  $3\sigma$ allowed regions of $M_H$, including all errors, are 
$[114,\,145]\, \rm GeV$ and $[[113,\,168]\;{\rm and}\;[180,225]]\, \rm GeV$, respectively. 
The inclusion of the direct Higgs search results from LEP leads to a strong rise of the 
$\Delta\chi^2$ curve below $M_H=115\,\rm GeV$. The data points from the searches at the 
Tevatron, available in the range $110\, {\rm GeV} < M_H < 200\,{\rm GeV}$ increases the 
$\Delta\chi^2$ estimator for Higgs masses above $140\,{\rm GeV}$ beyond that obtained 
from the {\em standard fit}.

The strong coupling at the $Z$-mass scale is determined by the {\em complete fit} with
$\alpha_S(M_Z^2)=0.1193^{\,+0.0028}_{\,-0.0027} \pm 0.0001$ where the first error is 
experimental and the second due to the truncation of the perturbative QCD series.  

Figure~\ref{fig:chi2lep} (right) compares the direct measurements of $M_W$ and $m_t$, 
shown by the shaded/green $1\sigma$ bands, with the 68\%,  95\% and 99\%~CL obtained 
for three sets of fits: the largest/blue (narrower/yellow) allowed regions are derived 
from the {\em standard fit} ({\em complete fit}) excluding the measured values in the 
fits. The inclusion of the LEP and Tevatron Higgs searches significantly impacts the 
constraints obtained. Good agreement between direct measurements and indirect fit results
is observed. The third set of fits (narrowest/green) results from the {\em complete fit} 
including the measured values. Hence it uses all available information and leads to 
the narrowest allowed region.

The $p$-value of the global SM fit, quantifying the probability of wrongly rejecting 
the SM hypothesis, has been evaluated by means of toy MC experiments. 
For each MC experiment, the {\em complete fit} is performed yielding the $\chi^2_{\rm min}$ 
distribution shown by the light shaded histogram in Fig.~\ref{fig:2hdm-combined} (left).
The monotonously decreasing curves give the $p$-value of the SM fit as a function of 
$\chi^2_{\rm min}$ obtained by integrating the distribution between $\chi^2_{\rm min}$ 
and infinity. The value of the global SM fit is given by 
$p\mbox{-value\,(data$|$SM)}\ = \ 0.22\pm0.01_{\,-0.02}\;$, where the first error is 
statistical and the second accounts for the shift resulting from theoretical uncertainties.

\section{CONSTRAINTS IN THE 2HDM}

As an example for a study beyond the SM we investigate models with an extended Higgs sector 
of two doublets. In the Type-II 2HDM, we constrain the mass of the charged Higgs and the 
ratio of the vacuum expectation values of the two Higgs doublets using current measurements 
of observables from the $B$ and $K$ physics sectors and their most recent theoretical 2HDM 
predictions, namely ${R_b^0}$~\cite{zsummary,haber}, the branching ratio (BR) of 
${B\to X_s\gamma}$~\cite{hfag,misiak}, the BR of leptonic decays of charged pseudoscalar 
mesons ($B\to\tau\nu$~\cite{chang,hou}, $B\to\mu\nu$~\cite{babar,hou} and 
$K\to\mu\nu$~\cite{flavianet}) and the BR of the semileptonic decay 
$B\to D\tau\nu$~\cite{dtaunu,kamenik}.

\begin{figure}[t]
  \centering
   \includegraphics[width=88mm]{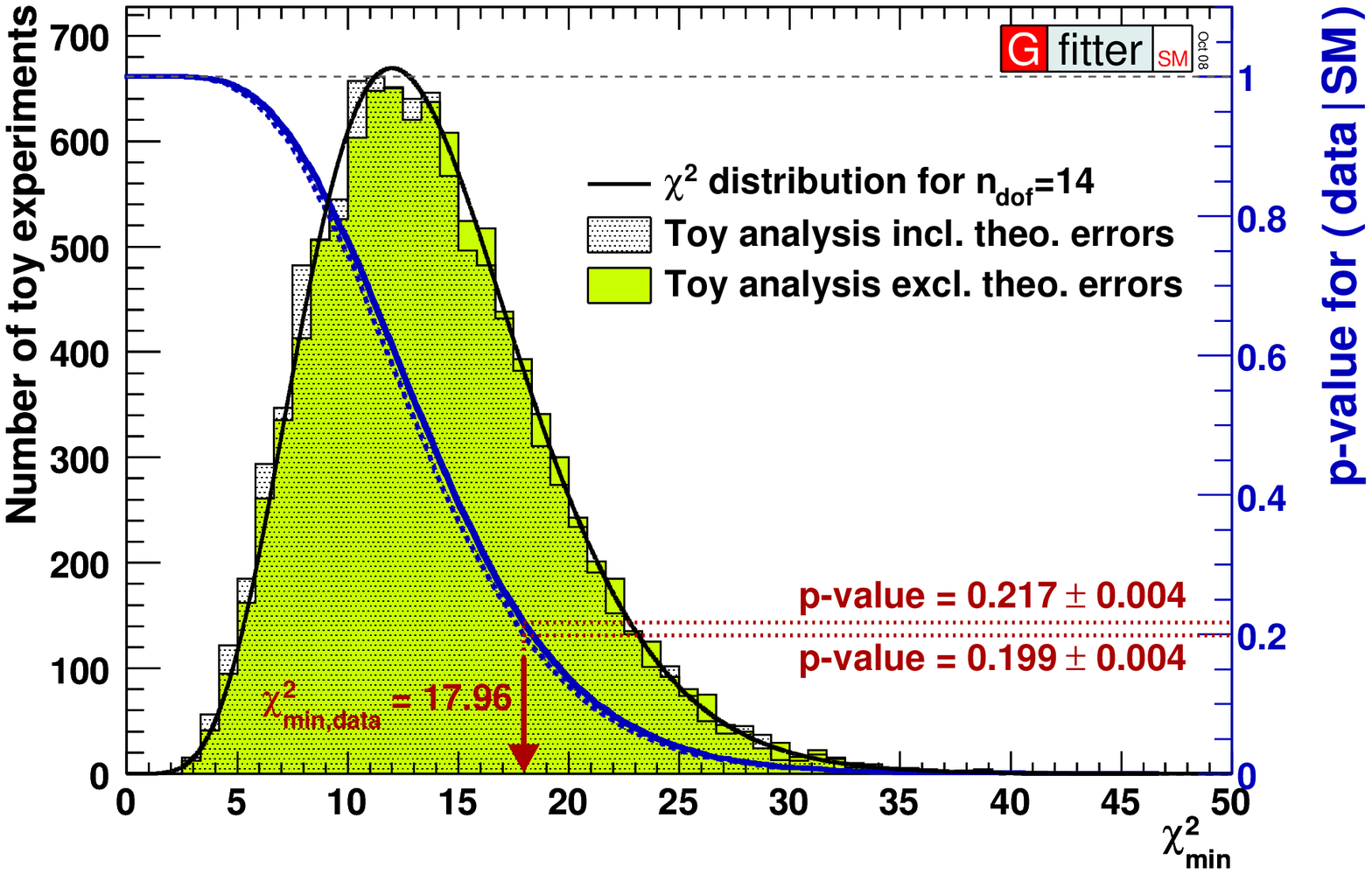}
   \includegraphics[width=88mm]{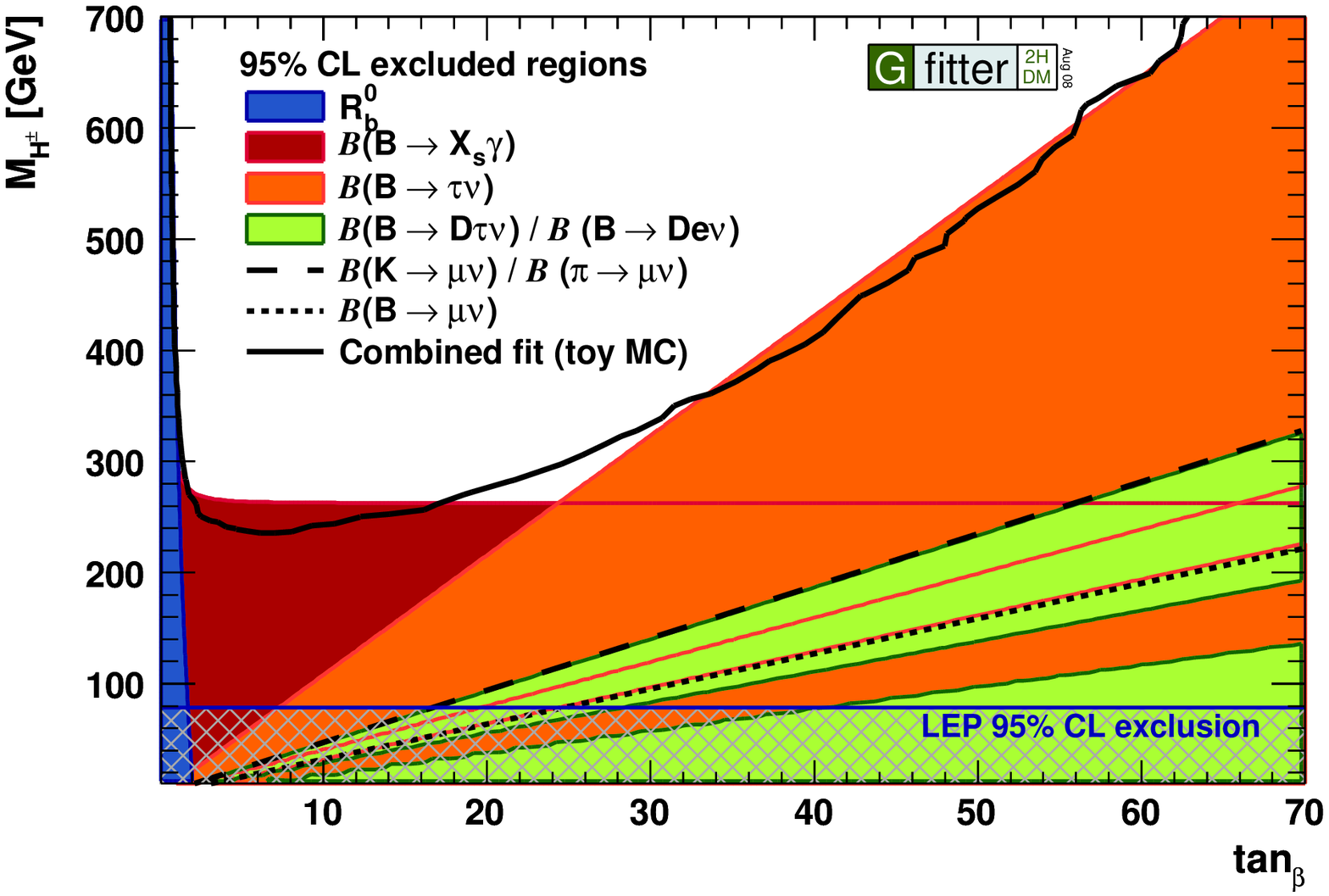}
   \caption{Left: Result of the MC toy analysis of the {\em complete fit} of the electroweak SM. 
     Shown are the $\chi^2_{\rm min}$ 
            distribution of a toy MC simulation (open histogram), the corresponding distribution 
            for a {\em complete fit} ignoring theoretical uncertainties (shaded/green histogram), 
            an ideal $\chi^2$ distribution assuming a Gaussian case with $n_{\rm dof}=14$ (black line) 
            and the $p$-value as a function of the $\chi^2_{\rm min}$ of the global fit; 
            Right: 95\% CL exclusion regions in the ($\tan \beta, M_{H^{\pm}}$) plane from 
            individual 2HDM constraints and the toy-MC-based result (hatched) from the combined 
            fit overlaid. }
   \label{fig:2hdm-combined}
\end{figure}

For each observable, individual constraints have been derived in the ($\tan\beta,M_{H^{\pm}}$) 
plane. Figure~\ref{fig:2hdm-combined} (right) displays the resulting 95\% {\em excluded} 
regions derived assuming Gaussian behaviour of the test statistics, and one degree of 
freedom. The figure shows that $R_b^0$ is mainly sensitive to $\tan \beta$ excluding small 
values. BR($B\to X_s\gamma$) is only sensitive to $\tan \beta$ for values below $\simeq$1. 
For larger values it provides an almost constant exclusion of a charged Higgs lighter than 
$\simeq$$260\,\rm GeV$. For all leptonic observables the 2HDM contribution can be either 
positive or negative since signed terms enter the prediction of the 
BRs resulting in a two-fold ambiguity in the $(\tan\beta,m_{H^\pm})$ space. 

In addition, we have performed a global fit combining the information
from all observables. 
For the CL calculation in the 2-dim plane we performed toy MC tests in each scan 
point which allows to avoid the problem of ambiguities in the effective number of 
degrees of freedom. The 95\% CL excluded region obtained are indicated in 
Fig.~\ref{fig:2hdm-combined} (right) by the area below the single solid line.
We can exclude a charged Higgs mass below $240\,\rm GeV$ 
independently of $\tan\beta$. This limit increases towards larger $\tan\beta$, 
e.g., $M_{H^{\pm}}<780\;\rm GeV$ are excluded for $\tan\beta=70$.

\end{document}